\documentclass[aip,
 apl,preprint,
 sd,%
 amsmath,amssymb,
 reprint,%
]{revtex4-1}
\usepackage{graphicx}
\usepackage{color}
\renewcommand{\bibnumfmt}[1]{[#1] } 
\usepackage{dcolumn}
\usepackage{bm}
\usepackage{float}

\begin{document}

\title{Control of light polarization by voltage tunable excitonic metasurfaces}

\author{S.V. Lobanov}
\affiliation{School of Physics and Astronomy, Cardiff University, Cardiff CF24 3AA, United Kingdom}
\affiliation{Skolkovo Institute of Science and Technology, Moscow 143026, Russia}

\author{N.A. Gippius}
\affiliation{Skolkovo Institute of Science and Technology, Moscow 143026, Russia}

\author{S.G. Tikhodeev}
\affiliation{A. M. Prokhorov General Physics Institute, Russian Academy of Sciences, Vavilova Street 38, Moscow 119991, Russia}
\affiliation{M. V. Lomonosov Moscow State University, Leninskie Gory 1, Moscow 119991, Russia}

\author{L.V. Butov}
\affiliation{Department of Physics, University of California at San Diego, La Jolla, California 92093-0319, USA}

\begin{abstract}
We propose a light emitting device with voltage controlled degree of linear polarization
of emission.
The device combines the ability of metasurfaces to control light with an energy-tunable light source based on indirect
excitons in coupled quantum well heterostructures.
\end{abstract}

\date{\today}

\maketitle

Plasmonic resonances in metal gratings on a dielectric substrate, or plasmonic metasurfaces, are now widely used  for various purposes
such as to enhance the efficiency of light emitting devices (LED)~\cite{Lozano2016}, to magnify the light absorption in
the photodetectors~\cite{Hetterich2007} and thin film solar cells~\cite{Panoiu2007,Pala2009}, to control the
intensity and directivity of light emission (as nanoantennas)~\cite{Neubrech2017}, to enhance the
magnetooptical effects~\cite{Belotelov2007,Chin2013,Floess2016}, the sensitivity of optical sensors for gas detecting, chemical
 and biosensing~\cite{Liu2010,Tittl2014,Jackman2017}, to magnify nonlinear optical effects such as 2nd and 3rd
 harmonic generation~\cite{Li2015}, in photochemistry~\cite{Zeng2016}.

\begin{figure}[ht]
\includegraphics{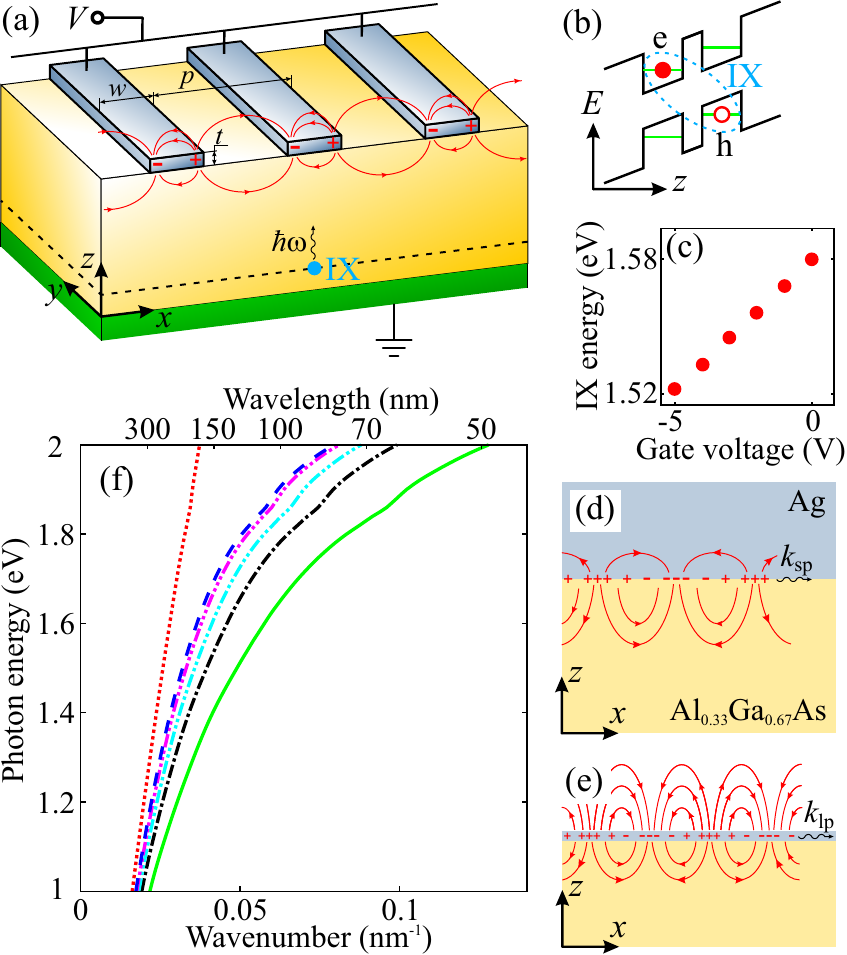} 
\caption{
(a)~Schematic diagram of the device. Coupled quantum well, CQW, (black dashed lines) is positioned within an undoped 1\,$\mu$m thick Al$_{0.33}$Ga$_{0.67}$As
layer (yellow) between a conducting $n^+$-GaAs layer serving as a homogeneous bottom electrode (green) and a striped top
electrode (silvery).
(b)~Energy band diagram of the CQW; e, electron; h, hole. The cyan dashed ellipse indicates an indirect exciton\,(IX).
(c)~Energy of IX as a function of applied voltage. The data is taken from Ref.\citenum{High2008}.
(d,e) Schematic of SPP propagation along a metal-dielectric boundary (d) and a thin metal film in asymmetric dielectric surrounding~(e).
(f)~Calculated dispersions of Ag/AlGaAs SPP (blue dashed line) and air/Ag/AlGaAs thin-layer SPP for silver film with thickness $t=10$\,nm
(green solid line), 15\,nm (black dash-dotted line), 20\,nm (cyan dash-double-dotted line), and 30\,nm (magenta dash-triple-dotted line).
Red dotted line shows the light cone in Al$_{0.33}$Ga$_{0.67}$As.
}
\end{figure}

In this letter we propose to combine the ability of plasmonic metasurfaces to control light with an energy-tunable light source based on indirect
excitons (IXs) for creating integrated optoelectronic devices where the polarization state of emitted light is controlled by voltage. IXs are composed
of electrons and holes in spatially separated quantum well layers (Fig.\,1b). Due to the IX built-in electric dipole moment $ed$, the energy of light
emitted by IXs is effectively controlled by voltage $\delta E = - edF_z$, where $d$ is the separation between the electron and hole layers
(for coupled quantum wells, CQWs, $d$ is close to the distance between the QW centers) and $F_z \propto V$ is an electric field perpendicular
to the QW plane. The IX energy control by voltage allowed creating a variety of excitonic devices, including traps and lattices for studying
basic phenomena in IXs, as well as excitonic transistors, routers, and photon storage devices, which form the potential for creating
excitonic signal processing devices and excitonic circuits~\cite{High2008,Winbow2007,Andreakou2014}.

Specifically, in this letter, we present a device demonstrating proof-of-principle for tunable excitonic metasurfaces, namely, a
light emitting device with voltage controlled degree of linear polarization (DLP) of emission made of plasmonic metal grating
deposited on top of an AlGaAs/GaAs CQW structure. The metal grating is dual-purpose.
On the one hand, it serves as the upper electrode to control the IX emission frequency. On the other hand,
it provides plasmonic resonances, to control the DLP
of IX photoluminescence.

The schematic diagram of the proposed device is shown in Fig.\,1a.
CQW (black dashed lines) is positioned within an undoped 1\,$\mu$m thick Al$_{0.33}$Ga$_{0.67}$As
layer between a conducting $n^+$-GaAs layer serving as a homogeneous bottom electrode and a top
electrode made of Ag nanowires grating. The nanowires have thickness\,$t$, width\,$w$ and the
grating period is\,$p$.

\begin{figure*}[ht]
\includegraphics{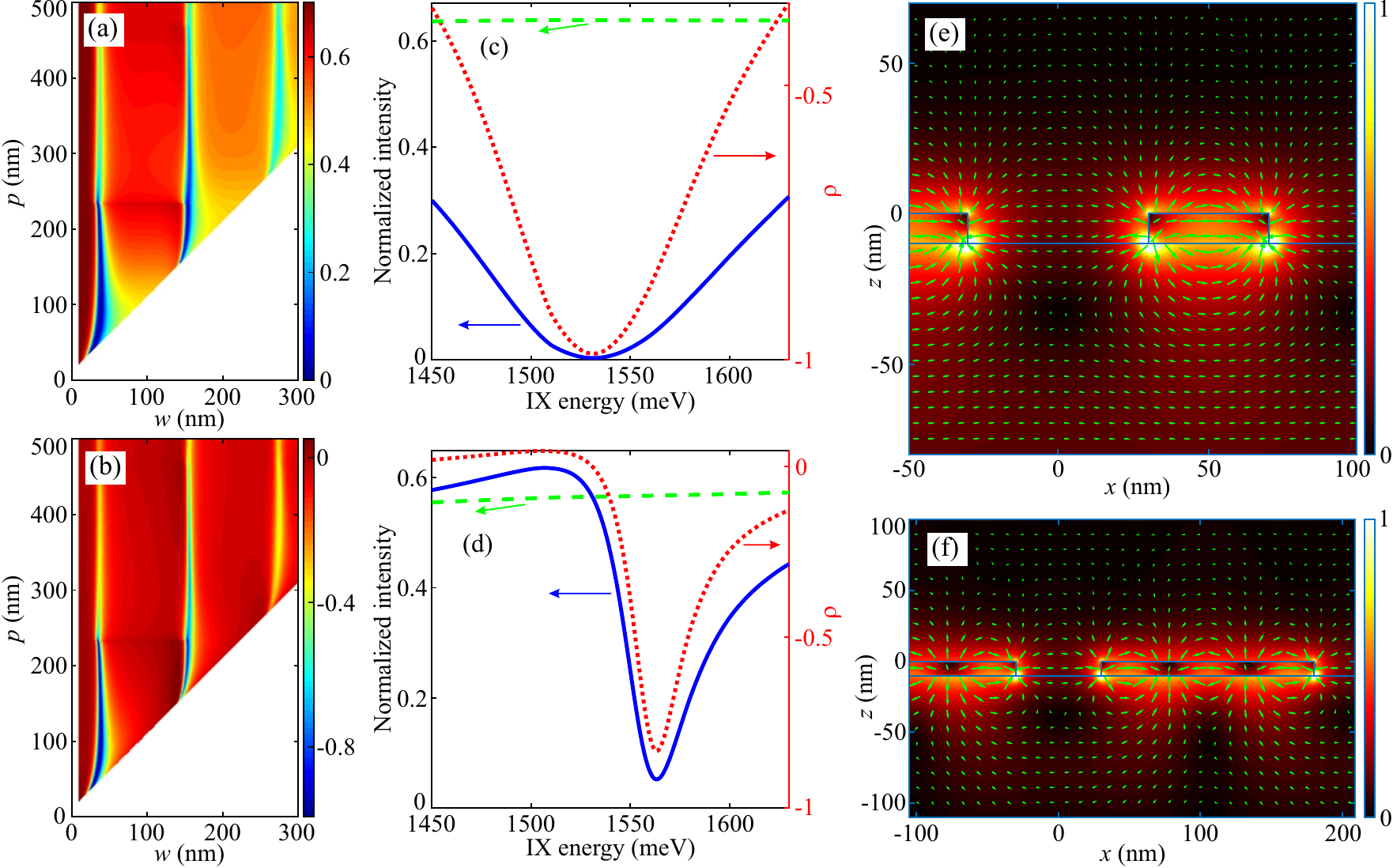} 
\caption{
(a)~Calculated dependence of $x$-linear polarized component of the IX device emission intensity $I_x$ along $z$-axis on the nanowires width $w$
and period $p$ for the photon energy $\hbar\omega=1.55$\,eV and thickness $t=10$~nm.
(b)~Respective degree of linear polarization\,$\rho$.
(c,d)~Calculated $x$- (blue solid line) and $y$- (green dashed line) linear polarized components of the emission spectra and respective degree
of linear polarization $\rho$ (red dotted line) for the structures with the stripe thickness $t$, width $w$, and period $p$, where
(c)\,$(t,w,p)=(10,40,100)$\,nm and (d)\,$(10,150,210)$\,nm.
The IX emission energy is controlled by voltage.
(e,f)~Calculated electric field distribution of the light propagating along $z$-axis with IX emission energy of (e)~$\hbar\omega=$1531\,meV and (f)\,1563\,meV for the structures with
(e)\,$(t,w,p)=(10,40,100)$\,nm and (f)~$(10,150,210)$\,nm.
Background color shows the absolute value of electric field.
}
\end{figure*}

The CQW band diagram is shown in Fig.\,1b; a typical dependence of the IX energy $E(V)$ on the applied voltage
is shown in Fig.\,1c\cite{High2008}.
Note however that $E(V)$ can be optimized for a tunable metasurface device by adjusting the CQW parameters
including the QW widths, width of the barrier between the QWs, and the well and barrier materials.
Figure 1c shows that changing the applied voltage allows to scan the energy of IX photoluminescence in the range of 1520--1580~meV.
If the metal grating is engineered in such a way that
one of its plasmon resonances is in the same energy range, it becomes possible to change the DLP of the device emission, due to pronounced changes of  $x$-polarized light transmission of the grating in the vicinity of the plasmonic resonances.

The characteristic energy dispersions of plasmons in metal graings can be simply understood as follows.
We start from a continuous metal film. Figure\,1d shows schematically the electric field of
a surface plasmon (SP) polariton (SPP), propagating along a boundary between a dielectric
and semi-infinite metal. The calculated SPP energy dispersion as a function
of wavenumber $k_\mathrm{sp}=2\pi/\lambda_\mathrm{sp}$ (where $\lambda_\mathrm{sp}$ is the SPP wavelength)
is shown for Ag/Al$_{0.33}$Ga$_{0.67}$As boundary by blue dashed line in Fig.\,1f. It is located
below the Al$_{0.33}$Ga$_{0.67}$As light cone (red dotted line in Fig.\,1f). The SPP dispersion curve
for air/Ag boundary (not shown) lies below the air light cone (also not shown), both are significantly
blue shifted due to a large refraction index of semicondicting AlGaAs.

If a semi-infinite metal is replaced by a metal film of finite thickness in asymmetric dielectric surrounding
(air and Al$_{0.33}$Ga$_{0.67}$As in our case), the plasmons propagating on the opposite sides of the metal film
interact with each other. The lower-energy SPP (localized predominantly on the AlGaAs side of the film)
repulses from the upper-energy SPP (localized predominantly on the air side). Resultantly, that yields a
red energy shift of the lower-energy SPP with the decrease of film thickness.
The calculated dispersions of the lower-energy SPP for silver film with thickness $t=10$\,nm (green solid line),
15\,nm (black dash-dotted line), 20\,nm (cyan dash-double-dotted line), and 30\,nm (magenta dash-triple-dotted line)
are shown in Fig.\,1f.
The dispersions are calculated neglecting absorption loses in silver and Al$_{0.33}$Ga$_{0.67}$As.

Note that
the polarization of all such SPPs is always magnetic field $\mathbf{H}\| y$ (see in Fig.\,1a), i. e., along the boundaries
and perpendicular to the SPP propagation direction $x$. As to the SSP electric field (shown schematically in Fig.\,1d-e),
there are both transverse and longitudinal components, $E_z$ and $E_x$, respectively.

If metal film is truncated from both sides into a single metal nanowire of width $w$, the localized
Mie plasmons arise due to SPP reflections back and forth, forming standing waves.
Formation of such localized plasmon requires the width of the wire to be roughly an integer number of thin-layer plasmon half
wavelength, i.e.
\begin{equation}\label{wm}
w_m=m\lambda_\mathrm{lp}/2,
\end{equation}
where $\lambda_\mathrm{lp}$ is wavelength of thin-layer plasmon and $m$ is a positive integer.
One can see from Fig.\,1f, that, e.g.,  the 10-nm-thick-layer plasmon wavelength is
$\lambda_\mathrm{lp}\approx 120$\,nm for the IX emission energy $\hbar\omega=1.55$\,eV.

Calculated dependence of $x$-linear polarized component of the IX device emission intensity $I_x$ along $z$-axis on the nanowires
width $w$ and period $p$ for the photon energy $\hbar\omega=1.55$\,eV and thickness $t=10$~nm is shown in Fig.\,2a.
In this and other figures, the emission intensity is normalized to the maximum emission intensity of an equivalent
IX in a homogeneous semiconductor Al$_{0.33}$Ga$_{0.67}$As. In order to calculate the emission intensity, we employ the
optical scattering matrix method~\cite{Tikhodeev2002,Lobanov2012} and the electrodynamical reciprocity principle
as described in Refs.\citenum{Maksimov2014,Lobanov2015,Lobanov2015a}, and use the dielectric
susceptibility of Ag from Ref.\citenum{Johnson1972} and Al$_{0.33}$Ga$_{0.67}$As from Ref.\citenum{Sopra1998}.

Three resonance branches are clearly seen in Fig.\,2a that correspond to excitation of localized SPPs.
These SPPs weakly depend on the period $p$. In a good qualitative agreement with
Eq.~(1) for odd $m$, these SPPs occur at nanowire widths
$w_1\approx40\,\mathrm{nm}\approx\lambda_\mathrm{lp}/2$, $w_3\approx150\,\mathrm{nm}\approx 3\lambda_\mathrm{lp}/2$,
and $w_5\approx270\,\mathrm{nm}\approx 5\lambda_\mathrm{lp}/2$.
The respective DLP $\rho=(I_x-I_y)/(I_x+I_y)$ is shown in Fig.\,2b.
Here, $I_y$ is $y$-linear polarized component of the IX device emission intensity.
One can see that DLP is varied in the range from about 0\% (unpolarized emission) to about -99\% for the first plasmon
branch ($w\approx40\,\mathrm{nm}$) and to about -85\% for the second branch ($w\approx150\,\mathrm{nm}$).

Calculated $x$- (blue solid line) and $y$- (green dashed line) linear polarized components of the IX device emission spectra and
respective DLP $\rho$ (red dotted line) for the structures with the nanowires thickness $t$, width $w$, and period $p$,
where $(t,w,p)=(10,40,100)$\,nm and $(10,150,210)$\,nm, are shown in Fig.\,2c and 2d, respectively.

In order to understand the effectiveness of the excitation of localized SPPs, we show in Figs.\,2e and 2f the calculated
electric field distribution of the light propagating along $z$-axis with photon energies 1531\,meV and \,1563\,meV,
for structures in Figs.\,2c and 2d, respectively.
The distributions have one and three anti-nodes, respectively, in agreement with the above discussion.
Note that SPPs with even $m$ have an even distribution of the electric field with respect to reflection in the vertical symmetry plane of the structure and, consequently, can not emit light along the $z$-axis for the structure of interest with spatially uniform excitation of CQW.

\begin{figure}[ht]
\includegraphics{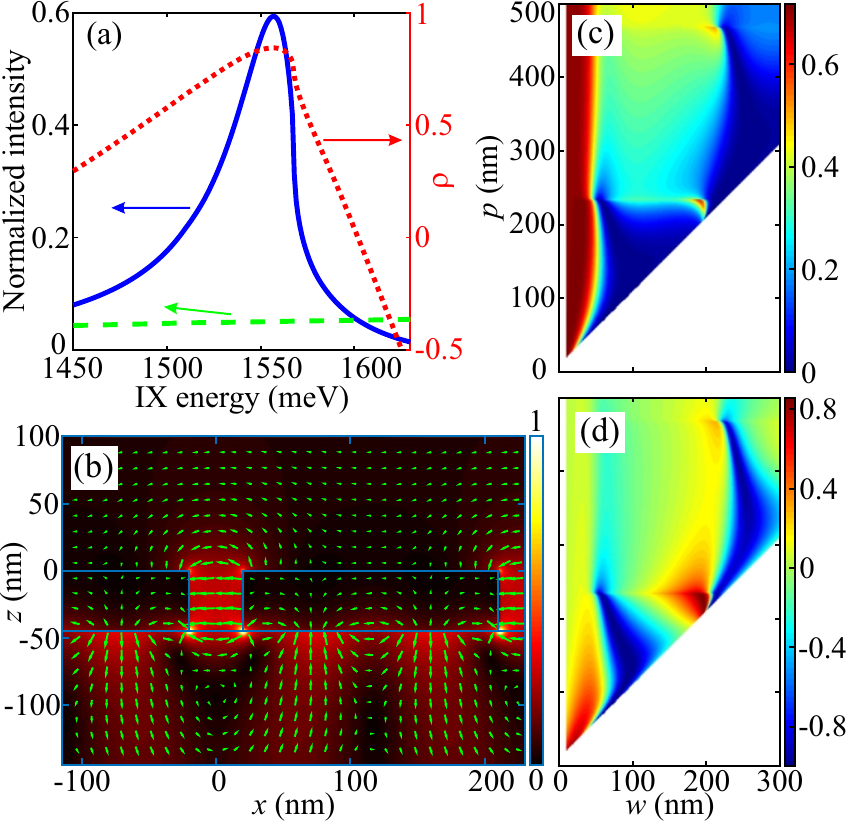}
\caption{(a) As Fig.\,2c, but for $(t,w,p)=(45,190,230)$\,nm.
(b)~As Fig.\,2e, but for $\hbar\omega=1557$\,meV and $(t,w,p)=(45,190,230)$\,nm.
(c)-(d)~As Figs.\,2a and 2b, but for $t=45$\,nm.
}
\end{figure}

In the described above cases of thin metal wires the transmissivity of the grating in both linear polarisations
along and perpendicular to the  wires is relatively high away from the SPP resonances. The emission of our device becomes
strongly linearly polarized along the wires ($\| y$, $\rho \sim -1$) at the SPP resonance frequency, because the grating becomes nearly
opaque for $x$-polarized light.

Using thick metal gratings, and the famous effect of the extraordinary optical transmission (EOT) through arrays of subwavelength
holes~\cite{Ebbesen1998,Treacy2002,GarciadeAbajo2007,GarciaVidal2010},
it is also possible to receive nearly completely $x$-polarized light emission of IXs near the SPP resonance frequency.

If thickness of a homogeneous metal film is larger than the skin depth of metal, which is about 20\,nm
for silver at $\hbar\omega=1.55$\,eV, interaction of surface plasmons propagating along opposite sides of the film is weak.
Therefore, one can consider them independently and use surface-plasmon dispersion curve (blue dashed line in Fig.\,1f) for analytical estimations.
One can see from Fig.\,1f that the wavelength of surface plasmon for photon energy $\hbar\omega=1.55$\,eV is $\lambda_\mathrm{sp}=184$\,nm.

If one makes narrow slits in a thick silver film, light can transmit through them.
The narrower are slits the smaller is transmission.
However, if the slits are ordered in a periodic array with period $p$, both light propagating along $z$-axis in Al$_{0.33}$Ga$_{0.67}$As
and surface plasmon propagating along $x$-axis along Ag--Al$_{0.33}$Ga$_{0.67}$As boundary undergo Bragg diffraction and
interact with electromagnetic waves, in-plane wavenumber of which differ by $2\pi m/p$, where $m$ is integer.
Taking into account that light propagating along $z$-axis has zero in-plane wavenumber, one conclude that this light can
excite surface plasmon with wavelength $\lambda_\mathrm{sp}$ if period of the grating is $p=m\lambda_\mathrm{sp}$,
where $m$ is a positive integer.
The surface plasmon in turn enhances electric field in the hole region, thus yielding
to EOT through the metal film with slits.

Calculated $x$- (blue solid line) and $y$- (green dashed line) linear polarized components of the IX device emission spectra and
respective DLP $\rho$ (red dotted line) for the structure with $(t,w,p)=(45,190,230)$\,nm is shown in Fig.\,3a.
One can see a narrow resonance with extraordinary emission of $x$-polarized light here.
Note, since the wavelengths of surface plasmon and light in Al$_{0.33}$Ga$_{0.67}$As are about the same for the photon
energy of 1.55\,eV (compare blue dashed and red dotted lines in Fig.\,1f), $x$-linear polarized component of the IX's
emission decreases rapidly from the right side of the resonance (blue line in Fig.\,3a) due to opening new diffraction channels in AlGaAs
(a Wood-Rayleigh anomaly).

The calculated electric field distribution of light propagating along $z$-axis with photon energy equal to the resonance energy
(1557\,meV) is shown in Fig.\,3b.
One can clearly see the excitation of SPP propagating along Ag/Al$_{0.33}$Ga$_{0.67}$As boundary and enhancement of electric
field in the hole region as it was discussed above.
Note also, surface plasmon form a standing wave or, in other words, propagates in both direction due to mirror symmetry of
the considered structure and normal light propagation.

Calculated dependence of $x$-linear polarized component of the IX's emission intensity $I_x$ along $z$-axis on the nanowires width $w$ and period $p$
for the photon energy $\hbar\omega=1.55$\,eV and stripe thickness $t=45$~nm is shown in Fig.\,3c.
A narrow resonance with extraordinary emission of $x$-polarized light occurs around the nanowires width $w=190$~nm and period $p=230$~nm.
These parameters correspond to the case of small slit between neighbouring nanowires $g = 40$\,nm.
Respective DLP $\rho$ is shown in Fig.\,3d.
One can see that DLP is varied in the range from about -100\% to about +85\%.

We also compared the device performance for various electrode materials. The estimates in the Supplementary Materials show
linear polarized components of the IX device emission spectra and respective DLP for voltage tunable excitonic
metasurfaces with gold gratings (Figs. S1-S3) operating in the same modes (see Figs. 2c, 2d, 3a) as the light emission device with silver gratings.
The DLP for these structures is varied in smaller range than for the silver ones because of larger absorptive losses of gold in the investigated frequency range.
Smaller DLP variations are estimated for metal nitride plasmonic materials~\cite{Naik2013} with larger losses such as titanium nitride,
see in Fig. S4.

To conclude, we propose voltage tunable excitonic metasurfaces --- integrated optoelectronic devices
with voltage control of the polarization state of light,
based on tunable emission of indirect excitons in coupled well semiconductor heterostructure
and plasmonic metasurface made of metal grating. Several modifications of such excitonic metasurfaces
are considered, which allow to tune the degree of linear polarization of emission in a
broad range from linearly polarized along grating to perpendicular to grating, depending on
the parameters of the gratings.

\begin{acknowledgments}
This work was supported by Russian Science Foundation
(Grant No. 16-12-10538), NSF Grant No. 1640173 and NERC, a subsidiary of SRC, through the SRC-NRI Center for Excitonic Devices.
The authors are thankful to T.~Weiss for fruitful discussions.
\end{acknowledgments}


\pagebreak
\widetext
\begin{center}
\textbf{\large Supplementary Materials: Control of light polarization by voltage tunable excitonic metasurfaces}
\end{center}
\setcounter{equation}{0}
\setcounter{figure}{0}
\setcounter{table}{0}
\setcounter{page}{1}
\makeatletter
\renewcommand{\theequation}{S\arabic{equation}}
\renewcommand{\thefigure}{S\arabic{figure}}
\renewcommand{\bibnumfmt}[1]{[S#1]}
\renewcommand{\citenumfont}[1]{S#1}


The Supplementary Materials presents calculated characteristics (linear polarized components of the emission spectra and respective DLP) of the proposed light emitting device with gold and titanium nitride gratings.
Figures S1 and S2 correspond to parameters of gold gratings where localized SPPs with $m=1$ and $3$, respectively, are excited.
Figures S3 and S4 correspond to parameters of EOT through arrays of subwavelength slits in gold and titanium nitride layers, respectively.

\begin{figure}[H]
\centering
\includegraphics{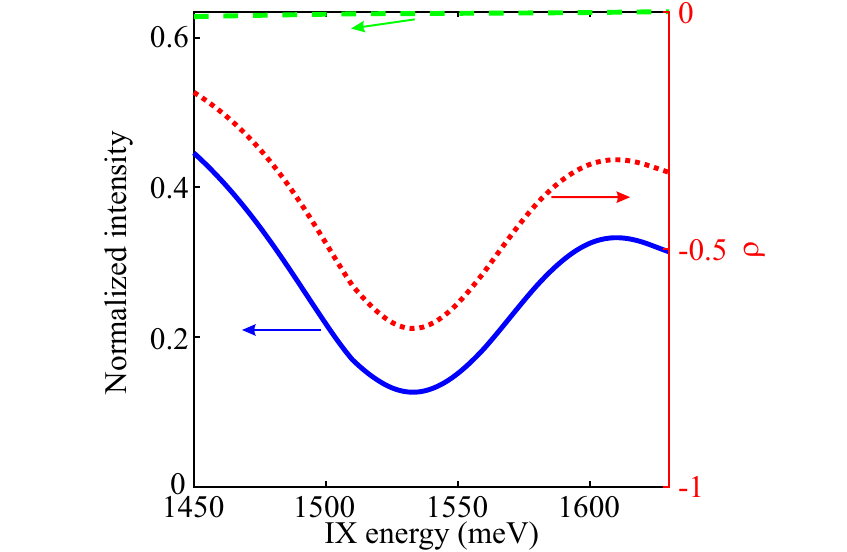}
\caption{As Fig.\,2c, but for gold gratings and $(t,w,p)=(10,23,50)$\,nm.
}
\end{figure}

\begin{figure}[H]
\centering
\includegraphics{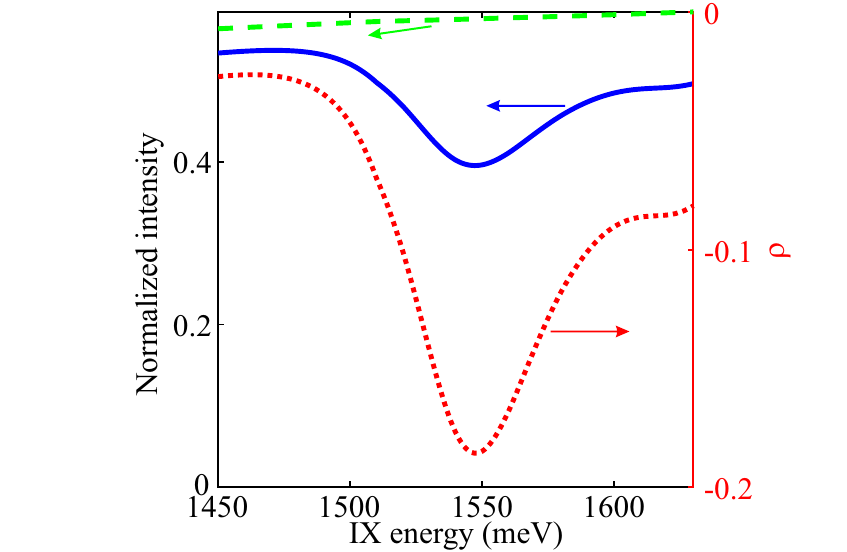}
\caption{As Fig.\,2c, but for gold gratings and $(t,w,p)=(10,120,160)$\,nm.
}
\end{figure}

\begin{figure}[H]
\centering
\includegraphics{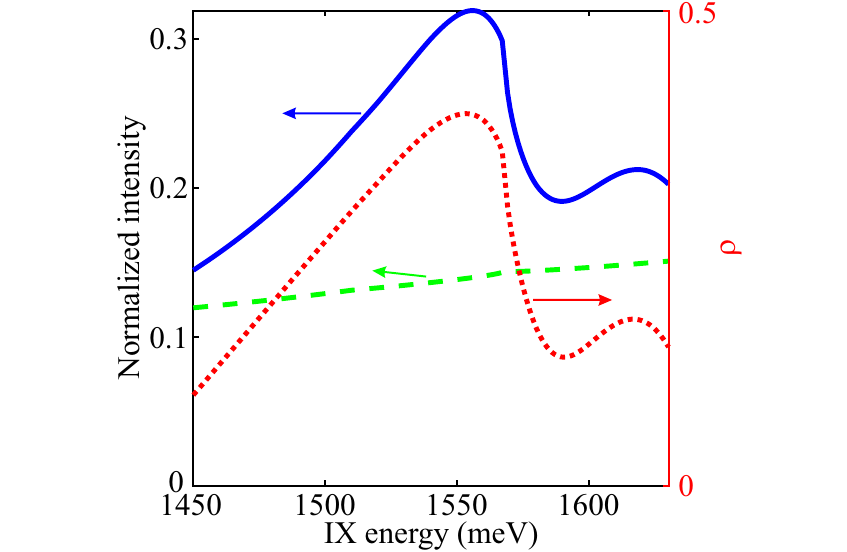}
\caption{As Fig.\,2c, but for gold gratings and $(t,w,p)=(45,160,230)$\,nm.
}
\end{figure}

\begin{figure}[H]
\centering
\includegraphics{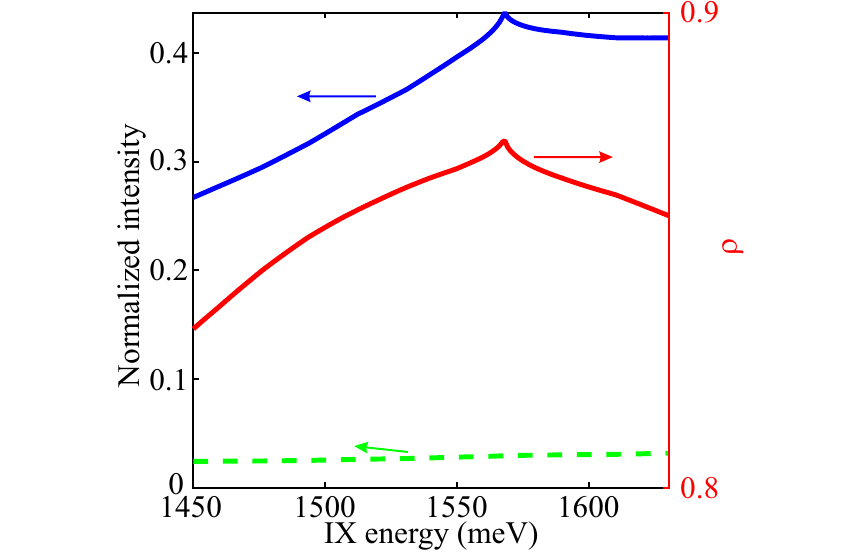}
\caption{As Fig.\,2c, but for titanium nitride gratings and $(t,w,p)=(100,150,230)$\,nm.
}
\end{figure}

\end{document}